# Enhanced sputter yields on ion irradiated Au nano particles: energy and size dependence

Henry Holland-Moritz[1], Sebastian Scheeler[2,3], Christoph Stanglmair[2,3], Claudia Pacholski[2,3] and Carsten Ronning[1]

[1] Institute for Solid State Physics, Friedrich Schiller University Jena, Max-Wien-Platz 1, D-07743 Jena, Germany
[2] Department of New Materials and Biosystems, Max Planck Institute for Intelligent Systems, Heisenbergstr. 3, D-70569 Stuttgart, Germany
[3] Insitute for Physical Chemistry, University of Heidelberg, Im Neuenheimer Feld 253, D-69120 Heidelberg, Germany

**Abstract**

Hexagonally arranged Au nano particles (NPs) exhibiting a broad size distribution ranging from 30 nm to 80 nm with a Gaussian shape were deposited on Si substrates and irradiated with $Ar^+$ and $Ga^+$ ions with various energies from 20 to 350 keV and 1 to 30 keV, respectively. The size and energy dependence of the sputter yield were measured using high resolution scanning electron microscopy image analysis. These results were compared to simulation results obtained by *iradina*, a Monte Carlo (MC) code, which takes the specifics of the nano geometry into account. The experimental obtained sputter yields are significantly higher compared to the calculated simulation results for both bulk and the nano geometry. The difference can be clearly attributed to thermally driven effects, which significantly influence the measured sputter yields.

# 1. Introduction

Miniaturization is a buzz word in developing new technologies. Nano science and nanotechnology are key disciplines for making devices and their elements smaller and more efficient. Nano particles (NPs) have already been introduced into everyday life [1]–[3]. However, still some work needs to be done to really make use of the mesoscopic properties of nano structures. These properties are mainly dominated by surface effects like surface reconstruction, depletion, Fermi level pinning and band bending. The reason is the increased surface-to-volume ratio of nano compared to bulk counterparts.

There are many ways to fabricate and grow nano structures, like physical vapor (PVD) or chemical vapor deposition (CVD) [4], [5]. Techniques like these work at or close to the thermal equilibrium, which limits i.e. solubility of doping atoms in a lattice or various other desired defects in the lattice. Ion implantation is a widely used subsequent approach to tune the properties of materials [6]–[10] and overcome the restrictions of the thermal equilibrium. An important effect, when it comes to ion irradiation, is sputtering. Sputtering is well understood theoretically by Sigmund's theory [11] as well as experimentally for bulk structures [12]–[15]. In nano structures, sputtering is enhanced due to the larger surface-to-volume ratio [16]. This enhancement plays a major role, i.e. when it comes to the doping of nano structures and the calculation of the doping concentration [17].

In this work, an approach of measuring the sputter yield of Au NPs, which are wet-chemically synthesized and subsequently deposited on a Si substrate by self-organization, is shown. We use high resolution SEM imaging to determine the sputter yields for $Ar^+$ and $Ga^+$ ions of various energies. Also, the size dependence of the sputter yield was investigated. These results are compared to the MC code *iradina* [18], which takes the specifics of nano structures into account.

# 2. Methods and Experimental

Gold nano particles were wet-chemically synthesized in water [19] or toluene [20] using seeding growth approaches. The resulting Au nano particles were equipped with a polystyrene shell by ligand exchange and subsequently spin-coated onto clean silicon substrates according to reference

[19], [20]. In figure 1 a representative scanning electron microscopy (SEM) image of a prepared Au nano particle array and the size distribution of the Au nano particles determined by transmission electron microscopy (TEM) image analysis are displayed.

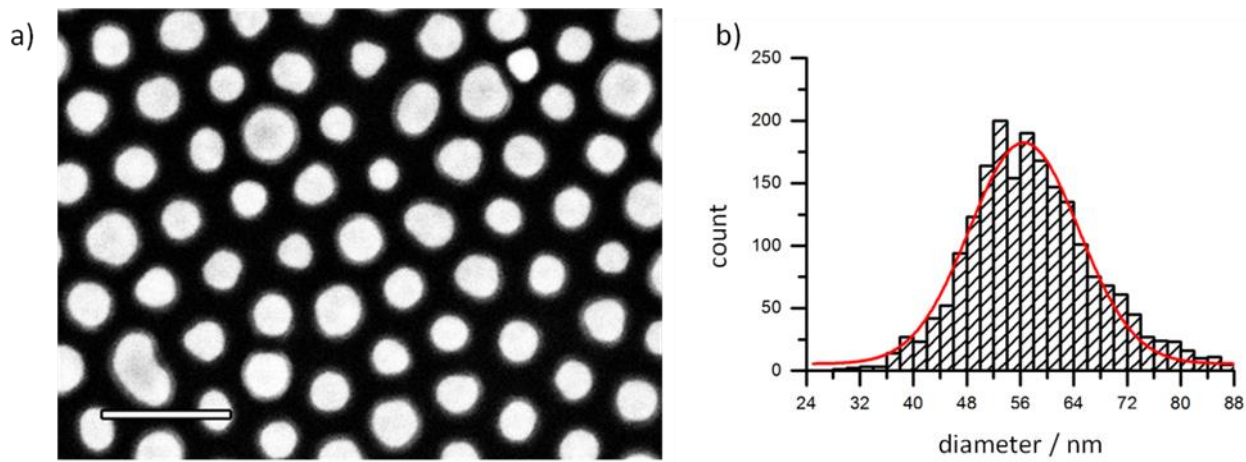

*Figure 1: Investigated Au nano particle arrays. a) Representative SEM image of a gold nanoparticle array fabricated by spin-coating of polystyrene-coated gold nanoparticles onto silicon substrates. Scale bar. 200 nm. b) Size distribution of wet-chemically synthesized Au nano particles determined by TEM.*

These samples were subsequently irradiated with $Ar^+$ or $Ga^+$ ions. The $Ar^+$ irradiations were performed using a general purpose ion implanter (High Voltage Engineering Europa). A FEI Helios i600 Nano Lab focused ion beam system (FIB) was used for the experiments with $Ga^+$ ions. In order to determine the energy dependence of the sputter yield, the samples were irradiated by $Ar^+$ ions with various energies ranging from 20 to 350 keV; the $Ga^+$ ion energy ranged from 1 to 30 keV. The total ion fluence for each sample was set to $3 \cdot 10^{15}$ cm$^{-2}$ and the ion beam current was kept constant at 0.5 µA and 0.1 µA for $Ar^+$ and $Ga^+$ ions, respectively. In order to evaluate the diameter dependence of the sputter yield, the ion energy was kept constant at 95 keV and 25 keV for $Ar^+$ and $Ga^+$, respectively. The ion fluencies for the $Ar^+$ and $Ga^+$ irradiations were chosen as $2 \cdot 10^{15}$ cm$^{-2}$ and $3 \cdot 10^{15}$ cm$^{-2}$, respectively. High resolution scanning electron microscopy (SEM) images of the samples were taken before and after each irradiation step. The position on the sample for image recording was chosen randomly for the energy dependent measurements. For the diameter dependence, the samples were marked/grooved with the $Ga^+$ beam of the FIB system in order to find the exact same position after the irradiation again. As a rough approximation, the NPs were assumed to be spherical.

Per sample and ion energy, about 1500 up to 3500 NPs were evaluated. The cross section area of the NPs were determined by ImageJ in order to determine the sputter yield [21]. An ellipse was automatically fitted to all NPs, if they showed a certain threshold circularity of 0.3. If a NP did not show this circularity, it was not evaluated in the further process. The measured areas before and after the irradiation were used to calculate the sputter yield using

$$Y = \frac{\rho_{at} \cdot \Delta V}{\Phi \cdot A_0} = \frac{4\rho_{at}}{\sqrt{\pi}\Phi \cdot A_0} \cdot (A_0^{3/2} - A_i^{3/2}),$$

where $Y$ is the sputter yield, $\rho_{at}$ is the target materials atomic density, $\Delta V$ the volume change due to the irradiation, $\Phi$ the ion fluence, and $A_0$ and $A_i$ the average cross section of the NPs before and after the irradiation, respectively. For the ion energy dependent experiments, the mean value of the NP areas were used for calculation, while for the diameter dependence the areas of exact the same NPs were taken into account.

Monte Carlo (MC) simulations were performed with the code *iradina* [18] in order to calculate the sputter yields and predict the volume decrease of the NPs. This code is based on *corteo* [22] and takes the specifics of the nano geometry into account. The binary collision approximation (BCA) is used and freestanding NPs are simulated. Also, the simulation takes place at virtually 0 K. Ten thousand ions were simulated for each simulation. The used material parameters for Au were: a displacement energy of 25 eV, a lattice binding energy of 2 eV and a surface binding energy of 3 eV [23]. The structure's surface of the NP was described analytically by a sphere. The simulation for the "nano geometry" was performed for a freestanding spherical Au NP with a diameter of 50 nm irradiated with Ar$^+$ ions of various energies. For diameter dependent simulations, the ion energy was set constant at 95 keV and 25 keV for Ar$^+$ and Ga$^+$, respectively, and the particle diameter was varied. The "bulk geometry" was described by an Au cuboid with periodic boundary conditions in the directions perpendicular to the beam direction. Here, only the ion energy was varied.

**3. Results and discussion**

**3.1 Energy dependence**

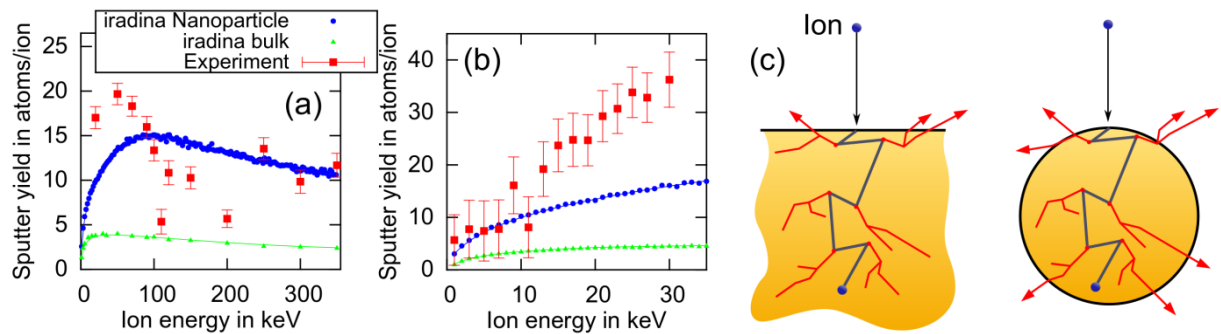

*Figure 2:* Energy dependence of the sputter yield for irradiated Au nano particles using $Ar^+$ (a) and $Ga^+$ (b) ions. The experimental results for $Ar^+$ ions show a large variation for ion energies larger 100 keV, but are higher than the simulated bulk yields in any case. The error bars of the experimental results represent the statistical errors of the mean value calculated from the standard deviation. (c) Schematic figure of geometries used in the MC simulations for the sputter yield: bulk (left) and nano geometry (right).

Figure 2 (a) shows the determined sputter yields for the $Ar^+$ irradiation of Au particles with an average diameter of 50 nm as a function of ion energy. The red data points represent the experimental results, while the blue and green dots denote the MC simulation results by *iradina* for "nano" and "bulk" geometry. The calculated sputter yield shows a steep increase with increasing energy for both geometries: bulk and NPs. The bulk sputter yield reaches its maximum of about 6 atoms/ion at an ion energy around 20 keV and then decreases with increasing energy, which is the typical energy dependence observed for flat surfaces and in agreement with reference [24], [25]. The MC simulation for the NPs shows an increase of the sputter yield up to 15 atoms/ion at an ion energy of 100 keV. At this ion energy, the ion range is comparable to the NP size of around 50 nm. One can notice that the MC simulation shows a considerably higher sputter yield for the spherical nano structure than for bulk geometry. The reason for the increased sputtering on nano structures is their increased surface to volume ratio. Figure 2 (c) illustrates this situation: the collision cascade is mainly inside the target for the bulk situation, and only a few recoils have a moving direction towards the surface and enough energy to overcome the surface binding energy to leave the target. The same ion trajectory is also shown for an ion hitting a nano structure with spherical cross section. Here, the damage cascade fills almost the whole NP, depending on the ion energy and the combination of target element and ion species. The cascade has more intersection points with the structure's surface, which leads to an increased number of sputtered atoms.

The experimental values, which are shown as red squares in figure 2(a), follow the given trend by the MC simulations of the "nano geometry" for small ion energies up to 100 keV. However, there is a notable shift of the maximum value of experimental sputter yield, which is at about 50 keV instead of ~ 100 keV. The sputter yield for the experimental results in this regime is also higher compared to the simulation by about 5 atoms/ion. For energies in the regime above 100 keV, the experimental values show a large variation. The value at 110 keV is the smallest value measured by the experiment, for energies larger than 250 keV, the variation decreases and the experimentally obtained values approach the simulation results for higher ion energies. In any case, the experimental results are larger than the simulated *iradina* "bulk" values. The discrepancy of simulation and experiment arises from several reasons, which will be explicitly addressed in the following. As a first reason, one can consider that the basic assumptions of the MC code are not accurate. In our simulation, a single, freestanding NP is represented, which neglects ion-substrate interactions as well as interactions between NP and substrate [26]. Additionally, *iradina* uses the binary collision approximation, which allows only the ejection of single atoms. However, the majority of sputtered atoms is ejected as clusters composed of up to thousands of atoms, as molecular dynamic (MD) simulations indicate for such an situation with a high mass target [16]. Furthermore, the MC code *iradina* calculates all processes at 0 K, which is obviously not true for the experiment. On the experimental side, the ion beam introduces thermal energy to the NPs, which leads to a considerably high amount of thermally evaporated atoms, which is discussed in detail in [16].

Some of the above mentioned effects, which lead to the strong fluctuations above ion energies of 100 keV, can be clearly seen in a representative SEM image of an irradiated sample shown in figure 3 (a). Firstly, one can see many small dots around the gold NPs, which are Au atoms/cluster sputtered in forward direction from the NPs and are redeposited at the sample surface. Secondly, it is noticeable, that different NPs reacted quite differently to the ion bombardment even though they were prior irradiation almost identical (compare figure 1). Some NPs, such as the one marked with #1, did not change their shape at all, while other NPs completely vanished, such as NP #2. Many other NPs, like #3, changed their shape dramatically and seemed to sink into the Si substrate. A possible explanation for the latter two effects is the large amount of thermal energy transferred to the NPs by the ion beam. The NPs have a small volume, a high surface-to-volume ratio and have a low thermal contact/conductivity to the Si substrate. The introduced ion energy ends up into phonons and simple thermal energy, which cannot dissipate quickly. While one ion heats the NP, another ion impinges while the temperature is still high, and heats the NP even more. Let assume, that one ion transfers its 100 keV as thermal energy to an Au NP with a diameter of 30 nm containing about $8.4 \cdot 10^5$ atoms. This leads to about 0.1 eV/atom and according to $E = k_B \cdot T$, which would result into a temperature increase of about 1000 °C, if any thermal losses like radiation, heat conduction and evaporation of atoms are neglected. Note, this simple estimation does not account for the extreme non-equilibrium process of ion irradiation. However, especially small NPs are affected by such thermal effects compared to larger ones with the same incident ion energy. In reality the temperature is of course smaller, but due to consecutive ion impacts, the melting and even the boiling point can be locally reached pretty fast. Some further effects, which can be attributed to thermally driven evaporation of material can be seen at NP #3 in figure 3 (a). This NP shows so called fingers, bridges and slingshots, which arise due to ejected heated and molten material. For bulk, similar effects have been shown by Nordlund et al. [27]. All these thermal effects, which are evident from the SEM figure 3(a), are not considered in MC codes at all. The pronounced effect on small NPs can be seen in figure 3 (b). Here, the size distributions of NPs are shown before and after the Ar$^+$ irradiation with 100 keV (left hand side) and 110 keV (right hand side). One can clearly see that the Gaussian shape is conserved in both cases. The mean value of the distribution shifts slightly towards smaller NPs for both ion energies, of course, due to sputtering. However, when we compare smaller NPs, we observe that the amount of NPs smaller than 40 nm is higher for 100 keV ions than after irradiation with 110 keV ions. There are two reasons for this: on the one hand, some small NPs simply vanish. On the other hand, some of these NPs weren't taken into account by the image analysis software (see above), because with increased deformation the NPs show a smaller circularity and are ignored by the software. This moreover leads to an apparent increase of the mean diameter for the nano particles after the irradiation, which causes a smaller sputter yield for the 110 keV experiment. For the ion energy of 110 keV and energies larger than 200 keV, more of the smaller NPs survive, which leads again to a larger sputter yield. This higher survival rate can be explained by the higher ion range in the material. Ions penetrate simply through the NPs and deposit less energy. For example, at 250 keV ion energy, an $Ar^+$ ion has an average ion range of 74 nm in Au, according to *TRIM* (Transport Of Ions in Matter) [28]. At this energy, most of the NPs with a diameter of 50 nm or less can be expected to be percolated by the ions.

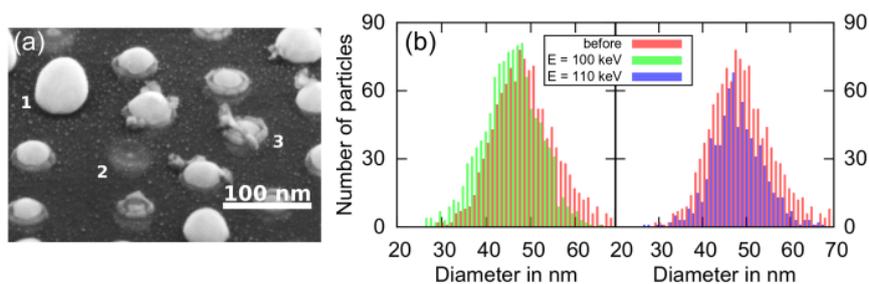

*Figure 3:* (a) Representative SEM image of Au NPs after ion irradiation with 110 keV $Ar^+$ ions, using a fluence of $3 \cdot 10^{15} cm^{-2}$. The image was gathered under an angle of 52°. (b) Size distribution of NPs before (red) and after 100 keV (left hand side, green) and 110 keV (right hand side, blue) $Ar^+$ ions.

For high ion energies at which ions can penetrate through the complete NP, ion beam mixing starts at the NP-substrate interface. Due to the elevated temperatures, Au silicate is formed [29], [30], which leads to preferential sputtering of silicon from this alloy. This reduces the sputter yield of gold and can be considered as one more reason for the lower sputter yields at higher energies. Such a forming of an alloy can be seen at NP #2 in figure 3 (a). At the position where the Au NP once was, the substrate appears brighter than in the surrounding due to the mass contrast in SEM imaging. Most likely this is gold silicate, which remains in the substrate. The NP on top got sputtered away.

Another possible explanation for the large variation of the experimentally obtained sputter yield is given by the fact, that for all ion energies the same ion beam current was used. Different heating rates due to the different input powers (ion energy × ion flux) of the material for different energies are the consequences. Therefore, smaller NPs are affected in a different amount in different energy regimes, which causes a systematic variation of the results.

In figure 2 (b) the experimental sputter yields of Au nano particles are shown after ion irradiation with low energetic $Ga^+$ ions together with respective MC simulations as a function of ion energy. Again, these data points show the experimental results determined by mean values of the NP diameters. Up to an ion energy of 12 keV, the measured sputter yields are comparable to the MC simulations, except the value at an ion energy of 10 keV. With increasing ion energy, the experimentally obtained sputter yield values show a steep increase, at the ion energy of 30 keV the experimental value is about 4 times higher than the value obtained by *iradina*. Here, most of the experimental sputter yields are larger than the simulated bulk yields. In these experiments, no variation is observed. Most of the experimental effects described above, which may result into a deviation, are not applicable in these experiments, because the ion energies were quite low compared to the $Ar^+$ irradiations and the ion beam current was much lower, at 0.1 µA. This leads to lower input powers over the whole energy range and thermal driven effects are less pronounced. Furthermore, the ion range of $Ga^+$ is low for the investigated energies. The ion range in Au is 7.7 nm for 30 keV $Ga^+$ ions, according to *TRIM* [31]. Nevertheless, also here it was noticed that very small NPs vanished, albeit the effect was way less pronounced.

Overall, the higher experimental values compared to the simulation results can be explain by the fact that the MC code neglects thermal effects like energy deposition as heat. Also, thermal spike effects which lead to collective cluster sputtering are not considered in the simulation.

**3.2 Size dependence**

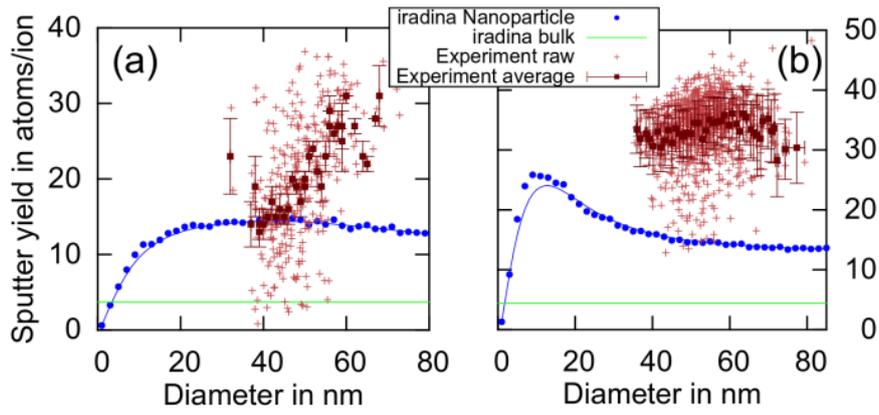

*Figure 4:* Size dependence of the sputter yield for spherical Au NPs for irradiation with (a) 95 keV Ar$^+$ ions and (b) 25 keV Ga$^+$ ions. The small crosses indicate the results for single NPs, the red squares the sputter yields averaged over intervals of 1 nm size.

In figure 4 (a), the sputter yields for both the simulation and experiments are shown for 95 keV Ar$^+$ irradiated Au NPs as a function of the size of the Au nano particles. According to the *iradina* simulations, the sputter yield is increasing until it reaches a maximum, (again) where the NP size is comparable to the ion range at this certain ion energy. In this increasing regime, the sputtering is mostly backward sputtering [32], since the ion range is large compared to the NP size and the ions interact mainly with the substrate. At the maximum, the damage cascade propagates over almost the whole NP and has the largest intersection area with the surface of the NP. The sputter yield decreases again for increasing diameters, the main fraction of the sputtering is due to side and backward sputtering [32]. The light red cross shaped data points show the experimental sputter yields for single NPs. One can see again a large variation of the data points. For example, NPs with a diameter around 39 nm show sputter yields between 1 and 30 atoms/ion. Since thermally driven effects are not pronounced for this situation, the reason must be different. One reasonable explanation is channeling. As discussed by Greaves et al. [16], the sputter yield strongly depends on the orientation of the NPs' lattice towards the ion beam. If the ion arrives in channeling direction, it passes a 20 nm Au NP without depositing much of its energy. But if the ion impacts in a random direction, the sputter yield is much higher [16]. Since the NPs are spin-coated onto the substrate, their orientation towards the ion beams is random. Some NPs are facing the ion beam in any lattice direction and some in random direction. This might explain the difference between the maxima and the minima sputter yields for certain NP sizes. Another attempt to explain this effect could be re-deposition of sputtered material on surrounding NPs. Smaller NPs have a larger sputter yield, larger NPs a smaller one. Since larger NPs cover a larger solid angle, the probability to collect sputtered atoms is larger. This might lead to a delayed sputtering for mid-range and larger NPs, somehow, a similar situation to Oswald ripening.

The variation was also observed in the experiments with $Ga^+$ ions, as shown in figure 4(b), but much less pronounced. The reason is that the ion range is much smaller compared to the size of the nano particles; whereas, for the above situation with 95 keV $Ar^+$ ions the average NP size matches the ion range. Here, the ion-solid interaction processes mainly occur in the upper hemisphere of the NPs with average and larger sizes. The channeling effects should be less important here.

The red data points in figure 4 show the average sputter yield values calculated from data points of NPs in intervals of 1 nm. For the $Ar^+$ irradiation, the average values for small sizes up to 45 nm fit the simulated data very well, for larger sizes the experimentally obtained sputter yield is increasing up to a factor of 3 compared to the MC simulation. By the models provided, this behavior cannot be explained. The sputter yields obtain for $Ga^+$ irradiations show also for the mean values less variation and smaller statistical errors. The values of the experimental sputter yields are about 3 times larger than the *iradina* results. As already mentioned above, MC calculations neglect thermal effects and cluster sputtering, which was shown by MD simulations to be the major driving force in the sputtering process [16]. The trend of the experimental results here proceeds parallel to the curve of the *iradina* results and decreases with larger NP diameter, as the model presented predict. This leads to the conclusion, that there is a systematic variation in the experiments for $Ar^+$ irradiated NPs, which tampers the results towards higher sputter yields for larger NPs and remains to be studied by future experiments.

The simulation results determined by *iradina* compared to the experimentally determined results are lower for Au compared to ion irradiated semiconductors [17]. With its higher density and mass, ion irradiated Au seems to show more pronounced thermally driven effects, especially "thermal spikes".

This enhances the sputter yield and shows the limit of the MC simulation.

## 4. Conclusions

An approach for the determination of the sputter yields of Au NPs was demonstrated, which is based on the automated analysis of a high number of high-resolution SEM images taken before and after ion irradiation. In parallel, simulations of sputtering were performed using *iradina*.

For the ion energy dependent measurements and a fixed Au average particle diameter of 50 nm, irradiations with $Ar^+$ showed a higher sputter yield for energies smaller than 100 keV compared to the simulation results. Thermally driven effects were clearly observed for higher ion energies, which lead to higher experimental sputter yields. Due to these effects, a large variation of the values was observed. In any case, the sputter yields were higher than the calculated bulk sputter yield. The reason for the deviation between simulation and experiment are the basic assumptions of the MC code such as BCA and neglecting thermally driven effects like thermal spikes and cluster ejection. Extreme deformations of the NPs by the thermally driven effects were only observed in the experiments carried out with $Ar^+$ ions. Also for $Ga^+$ ions the experimentally determined values were in most cases higher than the simulated sputter yields, except for small ion energies. The slope of the increase is steeper than the simulation predicts, since the experimental sputter yield is in most cases larger due to a larger amount of emitted atoms and clusters than the MC simulation predicts.

The measured sputter yields' size dependence for a fixed ion energy showed a large variation for the yields of individual NPs. This may be caused due to the different lattice orientations of various NPs towards the ion beam, as channeling will lower the sputter yield. Another effect might be re-deposition of sputtered material on surrounding NPs. The straggling for individual NPs was larger in the experiment with $Ar^+$ ions than with $Ga^+$ ions. The average values in intervals of 1 nm width showed less straggling. In the case of $Ar^+$ ions, the average sputter yields tended to increase for larger NPs, while the experimental $Ga^+$ yields were 2 times larger than the simulation results, but suited their trend.

The presented results show that sputtering is an important issue when irradiating nano structures with ion beams for doping, shaping or other modifications. Monte-Carlo simulations, like *iradina*, provide a good qualitative prediction of the amount of sputtered atoms in nano structures. But especially when irradiating metals, the sputter yield is underestimated.

**Acknowledgements**

We thank Patrick Hoffmann from the University of Jena for helping us in performing experiments at the accelerator. We also thank the BMBF for funding (BMBF project PhoNa, contract no. 03IS2101E & 03IS2101A) and the Max Planck Society.